\documentclass[sigconf]{acmart}

\AtBeginDocument{%
  \providecommand\BibTeX{{%
    \normalfont B\kern-0.5em{\scshape i\kern-0.25em b}\kern-0.8em\TeX}}}

\setcopyright{acmcopyright}
\copyrightyear{2018}
\acmYear{2018}
\acmDOI{XXXXXXX.XXXXXXX}

\acmConference[Conference acronym 'XX]{Make sure to enter the correct
  conference title from your rights confirmation emai}{June 03--05,
  2018}{Woodstock, NY}
\acmPrice{15.00}
\acmISBN{978-1-4503-XXXX-X/18/06}

\usepackage[english]{babel}
\usepackage{blindtext}
\usepackage{makecell}
\usepackage{indentfirst}
\usepackage{graphicx}
\usepackage{caption,subcaption}
\usepackage{multirow}
\usepackage{url}
\usepackage{algorithm}
\usepackage{algpseudocode}
\usepackage{dblfloatfix} 
\usepackage{listings}

\usepackage{soul}
\usepackage{kotex}

\newcommand*{\sh}[1]{\textcolor{blue}{#1}}
\newcommand*{\cmt}[1]{\textcolor{magenta}{#1}}

\author{
    {\rm Na Hyeon Park\textsuperscript{1}\textsuperscript{\textsection}} \hspace{1em} 
    {\rm Hanna Kim\textsuperscript{1}\textsuperscript{\textsection}} \hspace{1em} 
    {\rm Chanhee Lee\textsuperscript{2}} \hspace{1em} 
    {\rm Changhoon Yoon\textsuperscript{2}} \hspace{1em} \texorpdfstring{\\}
    {\rm Minsu Kim\textsuperscript{2}} \hspace{1em} 
    {\rm Seunghyeon Lee\textsuperscript{2}} \hspace{1em} 
    {\rm Youngjin jin\textsuperscript{1}} \hspace{1em}
    {\rm Seungwon Shin\textsuperscript{1}} \hspace{1em} \texorpdfstring{\\}
    \textsuperscript{1}KAIST, Daejeon, South Korea \texorpdfstring{\\}
    \textsuperscript{2}S2W Inc., Seongnam, South Korea\texorpdfstring{\\}
    \textsuperscript{1}\texorpdfstring{\texttt{\{julia19, gkssk3654, ijinjin,claude\}@kaist.ac.kr}\\}
    \textsuperscript{2}\texttt{\{leemember, cy, minsu, sl\}@s2w.inc}
} 

\begin{document}
\title{A Deep Dive into NFT Whales: A Longitudinal Study of the NFT Trading Ecosystem}


\renewcommand{\shortauthors}{Park and Kim, et al.}

\begin{abstract}

NFT (Non-fungible Token) has drastically increased in its size, accounting for over \$16.9B of total market capitalization. Despite the rapid growth of NFTs, this market has not been examined thoroughly from a financial perspective. In this paper, we conduct methodical analyses to identify NFT market movers who play a significant role in potentially manipulating and oscillating NFT values. We collect over 3.8M NFT transaction data from the Ethereum Blockchain from January 2021 to February 2022 to extract trading information in line with the NFT lifecycle: (i) mint, (ii) transfer/sale, and (iii) burn. Based on the size of held NFT values, we classify NFT traders into three groups (whales, dolphins, and minnows). In total, we analyze 430K traders from 91 different NFT collection sources. We find that the top 0.1\% of NFT traders (i.e., whales) drive the NFT market with consistent, high returns. We then identify and characterize the NFT whales' unique investment strategies (e.g., mint/sale patterns, wash trading) to empirically understand the whales in the NFT market for the first time. 

\end{abstract}

\maketitle

\begingroup\renewcommand\thefootnote{\textsection}
\footnotetext{Both authors contributed equally to this research.}
\endgroup



\received{20 February 2007}
\received[revised]{12 March 2009}
\received[accepted]{5 June 2009}

\section{Introduction}

Cryptocurrency, which has recorded a phenomenal growth in the past decade, has now become one of the most popular investment options. Successfully demonstrating the potential of its underlying blockchain technology as well as its enormous profitability, other blockchain-based digital assets have also emerged as alternative, life-changing investment opportunities. 

Following the huge success of cryptocurrencies, public attention toward NFTs has exploded. Many investors have traded NFT collectibles for financial gain, and eventually, the \emph{whales} appeared on the market. 
The term \emph{whale} has become a commonly accepted designation in the blockchain community. For example, in the cryptocurrency market, Satoshi Nakamoto and the Winkelvoss Brothers are often referred to as \emph{Bitcoin whales}. They own an exceptionally large amount of Bitcoins enough to impact the entire market as previously measured. ~\cite{bitcoin_whale, bitcoin_whale2, bitcoin_whale3}.

NFT whales can be easily monitored on commercial NFT analytics platforms~\cite{nftgo, nftbank} available today. For instance, NFTGo~\cite{nftgo}, one of the most popular NFT analytics platforms, evaluates the total value of the NFTs in each NFT holder's wallet and publishes a list of whales. The purpose of this feature is to provide useful investment information for potential NFT buyers who wish to follow the whales' investment pattern. As such, the movement of NFT whales, which may be potentially driving the market, may provide useful insights or lessons; however, neither whale-following investment strategy nor other aspects of NFT whales have been thoroughly studied.

To gain a deeper understanding of obscure NFT whales, we ask the following questions:
\begin{itemize}
    \item What are the characteristics of NFT whales?
    \item What are the investment strategies of NFT whales?
    \item Are NFT whales successful investors?
    \item Are NFT whales really powerful enough to impact the entire NFT market?
\end{itemize}

This paper dives deeply into the newly emerging NFT market to reveal and understand NFT whales for the first time. We collect the transaction logs from NFT smart contracts~\cite{event_logs} and align the data according to the NFT life-cycle: (i) mint, (ii) transfer/sale, and (iii) burn. Then, we estimate how much value has been transferred through NFT trades based on the market price at the time of each transaction. As a result, we are able to identify 430K traders involved in the top 91 NFT collections.

Below are the significant findings of this paper:
\begin{itemize}
  \item Most high-value NFTs are owned by NFT whales. 
  \item NFT whales rarely sell NFTs while eagerly buying the highest priced NFTs (top 1 percent)
  \item NFT whales are indeed the most profitable group.
  \item NFT whales take advantage of wash trading. 
\end{itemize}

The findings of this longitudinal measurement study have given us a lesson regarding the NFT market. Our study has shown that NFT whales have been powerful enough to manipulate the entire market. In other words, the NFT market itself is still immature and unstable. 
Furthermore, we also observe that the whales have been abusing weaknesses of the NFT market to inflate the value of the NFTs they own. In order to establish a mature and stable NFT market, the existing regulatory and legal environments must adapt to include NFT. It is also recommended to implement security countermeasures to protect the market from any abusive behaviors. 


\section{Background}
\label{s:background}
\noindent\textbf{Non-fungible token.} An \textit{NFT} is unique, irreplaceable, and its value is decided individually depending on demand. An NFT is a digital asset that represents objects like art, video, in-game entities, etc. NFTs are stored in blockchain, with Ethereum being the representative choice.
NFTs are categorized in a \textit{collection}, which refers to a set of NFTs. 
Commonly, NFTs in the same collection share common features. For example, one of most popular collection named \textit{Bored Ape Yacht Club (BAYC)} is a collection consisting of 10,000 ape tokens. 
The price of NFTs varies widely even within the same collection, although the range varies depending on the collection.


\noindent\textbf{NFT life-cycle.} Due to the anonymous nature of cryptocurrency wallet, we define a unique wallet address as a \textit{trader}. 
Here, we detail the procedures for how an NFT transaction operates, which consists of three steps: (i) mint, (ii) transfer/sale, and (iii) burn. 
\textit{Mint} refers to the process of offering newly created tokens to the public. Generally, NFT creators post the detailed schedule and price on various channels, such as discord, which is then conducted by a smart contract on a blockchain.
\textit{Transfer} is an act of simple value transfer which only passes ownership of NFT tokens to another wallet. \textit{Sale} is the process of transferring ownership of NFTs for a price, and is commonly held in NFT markets~\cite{opensea, looksrare}. 
\textit{Burn} refers to the method of removing ownership of NFT tokens on purpose. Once an NFT is burned, no one is able to gain its control.

\noindent\textbf{Classifying NFT traders.} There has been no official term for traders (individiuals or entities) that hold differing amounts of NFTs.
Thus, we consider each wallet as an individual \textit{trader} and classify the traders into three groups according to their holding values of NFTs, for every time range in our dataset:

\begin{itemize}
    \item \textbf{Whale} - top 0.1\% traders
    \item \textbf{Dolphin} - top 10\% traders excluding whales
    \item \textbf{Minnow} - all other traders excluding whales and dolphins (89.9\%)
\end{itemize}

To clearly identify the trader groups, we define \textit{holding value} to be used as a concrete criteria for our classification. 
The estimation of the current market price of NFTs may be imprecise due to the high volatility of the market value of NFTs compared to traditional economics. 
Instead, we believe that the latest trading price of each NFT could represent its value well. 
We define \textit{holding value of a trader} as the sum of the last traded price for each token held by the trader.

To focus on the top holding value traders, we refer to whales and dolphins collectively as \textit{holding value leaders}.
The number of traders in each group is shown in Table~\ref{t:data_collection} and is further described in Table~\ref{t:num_of_trader} of the Appendix.





\section{Tracing NFT Transactions on the blockchain}
\label{s:data}
In Ethereum, there are two types of transactions: \textit{external} and \textit{internal} transactions. 
Ether (ETH) transfers between users is recorded as external transactions.
External transactions have information such as receiver’s and sender’s addresses and transferred amount in ETH is recorded in the blockchain and readily available to anyone for reference.

On the other hand, transferring tokens, such as NFTs or fungible tokens, is a type of an internal transaction, which is not stored on the blockchain.
Instead, we can use the \textit{token transfer log}, which is recorded by token contracts when token transferring occurs. 
By collecting transfer logs from NFT contracts, we can trace how NFT transactions work in the NFT ecosystem.

We collect external transactions and token transfer logs from the Ethereum blockchain to track NFT ownership changes and subsequent payments. More details can be found in Section~\ref{sec:NFT_tx_collect}.

For the NFT ecosystem, 2021 marks the first year when the market began to grow rapidly with public attention~\cite{nft_boom1}, with NFT trading volumes showing an increase of 21,000\% from 2020~\cite{nft_boom2}. 
As a result, we focus on data with \texttt{block\_timestamp} from the first day of 2021 to February 28, 2022 (14 months). 
During this collection period, we obtain 3,838,587 transactions for over a million NFT items in total. Also, the unique number of accounts participating in a transaction at least once is 430.2K. Our data collection is summarized in Table~\ref{t:data_collection}.

\begin{table}[t]
\caption{Summary of data collection. The accounts are divided by whale, dolphin, minnow (February, 2022)}
\label{t:data_collection}
\centering
\footnotesize
\begin{tabular}{l l r}
    \toprule
    \multicolumn{2}{l}{\textbf{Type}} & \textbf{Collection} \\
    \midrule
    \multicolumn{2}{l}{NFT} & 1,129,6967 \\
    \multicolumn{2}{l}{Transaction} & 3,838,587 \\
    \multicolumn{2}{l}{Account} & 3,086,046  \\

    & Whale & 430  \\
    & Dolphin & 42,593  \\
    & Minnow & 387,204 \\ \cmidrule{2-3}
    & Total accounts & 430,277 \\
    \midrule
    \midrule
    \multicolumn{2}{l}{\textbf{Period}} &
    \multicolumn{1}{l}{January 1,2021 $\sim$ February 28, 2022}\\
    \bottomrule
\end{tabular}
\vspace{-10pt}
\end{table}

\section{Characteristics of NFT WHALES}
\label{s:basic}
In this section, we characterize NFT whales through a deep analysis of their behavior patterns such as unique trading methods for NFT items and portfolio management.

\begin{figure*}[t]
\vspace{0.1cm}
  \centering
  \includegraphics[width=0.8\linewidth]{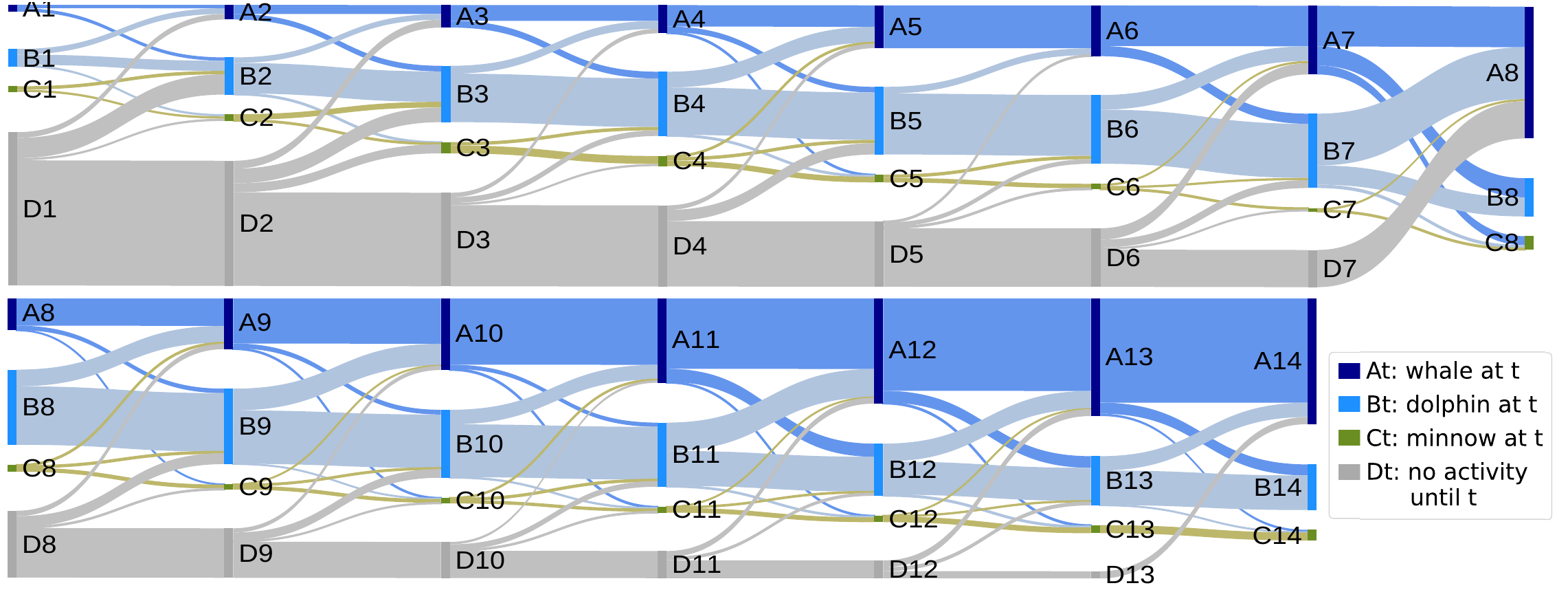}
  \caption{Changes in the group of traders in \textit{whale} from January 2021 (t = 1) to February 2022 (t = 14). The top and bottom only depict traders that have belonged to the \textit{whale} group at least once, until August and February 2022, respectively.
 Transitions from one color to another depict a quantity of traders with group changes. }
  \label{f:sankey}
\end{figure*}

\subsection{Changes in Whales' Composition}

The NFT market has changed dramatically over the past 14 months with the influx of new traders and the advent of various types of new collections. 
In this rapidly changing market, \textit{how do traders become whales}?
To answer this question, we study how traders move across between groups during 14 months, especially focusing on whales. Figure \ref{f:sankey} shows the changes in composition of the whale group until February 2022. 

In the beginning, since the size of whale group is small, the composition of the group is highly volatile as the market size increases. However, from June 2021, more than 80\% of whales remain in the next whale group, except in August 2021. August is when the most significant change in the whale group occurs. The surge of public interest at that time~\cite{nft_august} caused the largest number of new traders compared to the previous months. 
To understand how traders become whales and how they maintain their status, we investigate the characteristics according to the whales' origin as follows. The percentage of each group where the whales belonged to is summarized in Table \ref{t:across_users} of the Appendix.

\noindent{\textbf{Whales from \textit{whales}.}}
Whales maintain their status through various ways: buying, receiving, and minting. In the first half, more than half of whales buy NFTs in order to increase the holding value. However, in early 2022, whales who mainly receive tokens grow in number. Meanwhile, some whales actively mint NFTs. For example, one whale minted a tremendous number of tokens in several collections and sold them to dolphins and minnows. There also exist whales who were inactive for more than a few months. Their extremely expensive tokens let them stay in the whale group. \\
\noindent{\textbf{Whales from \textit{dolphins}.}}
As the size of the whale group increases, it is natural for dolphins to become whales. 
In August, when the market size grows most rapidly, dolphins take up 39\% of the new whale group. 
Dolphins raise their holding value in various ways like whales from \textit{whales}. 
However, in early 2022, the percentage of new whales from dolphins see a decrease to 15\%.


\noindent{\textbf{Whales from \textit{minnows}.}}
We rarely observe whales who formerly belonged to \textit{minnows}.
They increase their NFT holding value through buying or minting. 
They barely receive tokens from other traders, which indicates that they do not form particular relationships with other traders. 
This is completely different to first-time trading whales, which is described in detail below.

\noindent{\textbf{First-time trading whales.}}
There are a lot of traders who become whales as soon as they begin participating in the NFT market. When the size of the market increases rapidly, whales from this group sometimes outnumber the whales from the dolphin group. They usually become whales through buying or receiving, but many of them become whales only through the act of receiving tokens.
Usually, the tokens they receive are NFTs of popular collections with high trading volume (e.g. \textit{CrytoPunks}). Interestingly, most of the NFTs they received come from the former whales, implying a close relationship between them.\\
\noindent{\bf Findings and Insight.}
Although the NFT ecosystem is rapidly growing, whales maintain their place firmly. In addition, newly emerging whales are often associated with former whales. This clearly shows that it is hard for minnows to become whales.


\begin{figure}[ht]
  \centering
  \includegraphics[width=0.8\linewidth]{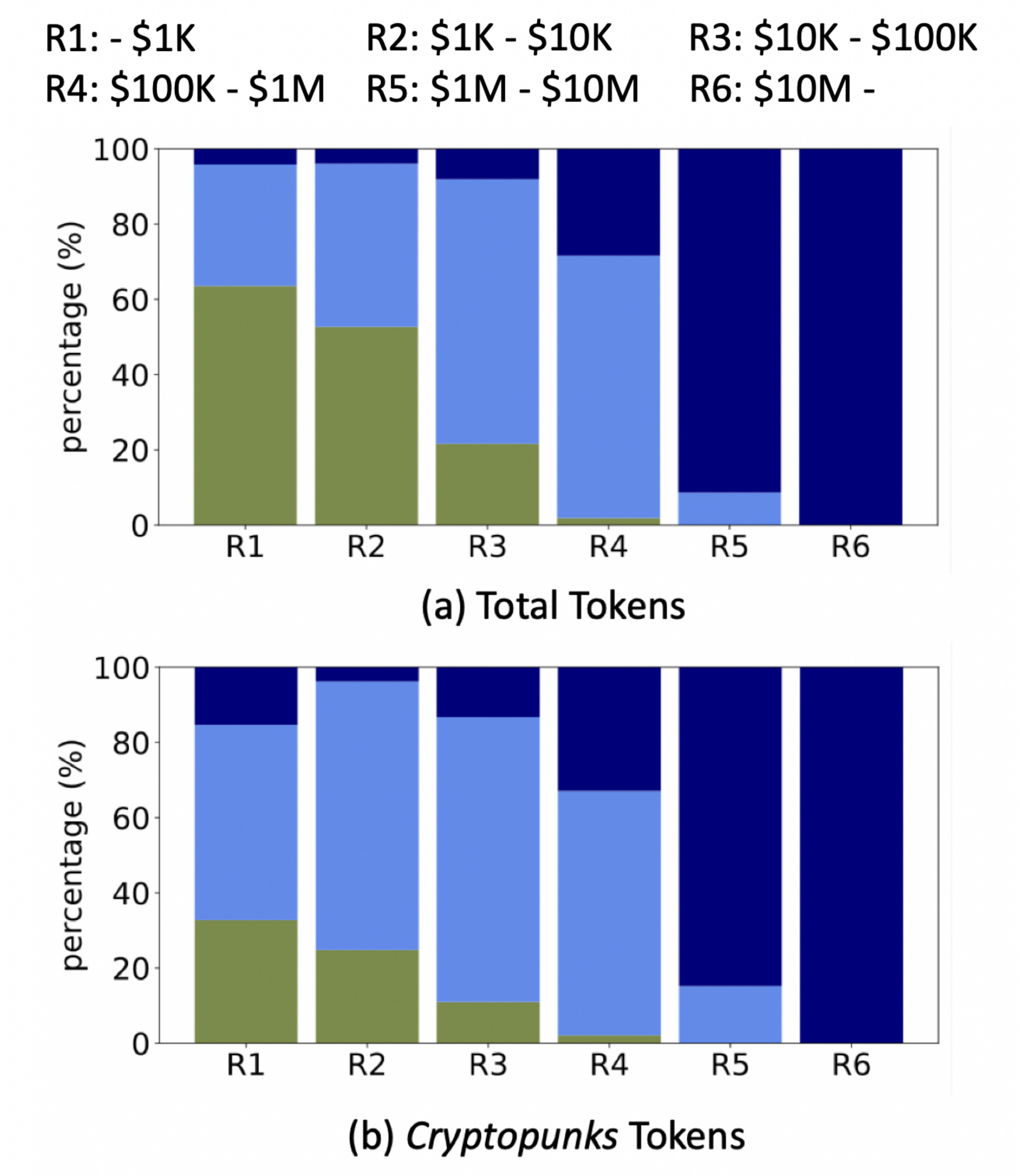}
  \caption{Price distribution of tokens held by each group. Dark-blue,
sky-blue, olive each represents holdings of whales, dolphins, and minnows.}
  \label{f:price_dist}
\end{figure}

\begin{figure*}[ht!]
  \centering
  \includegraphics[width=0.9\linewidth]{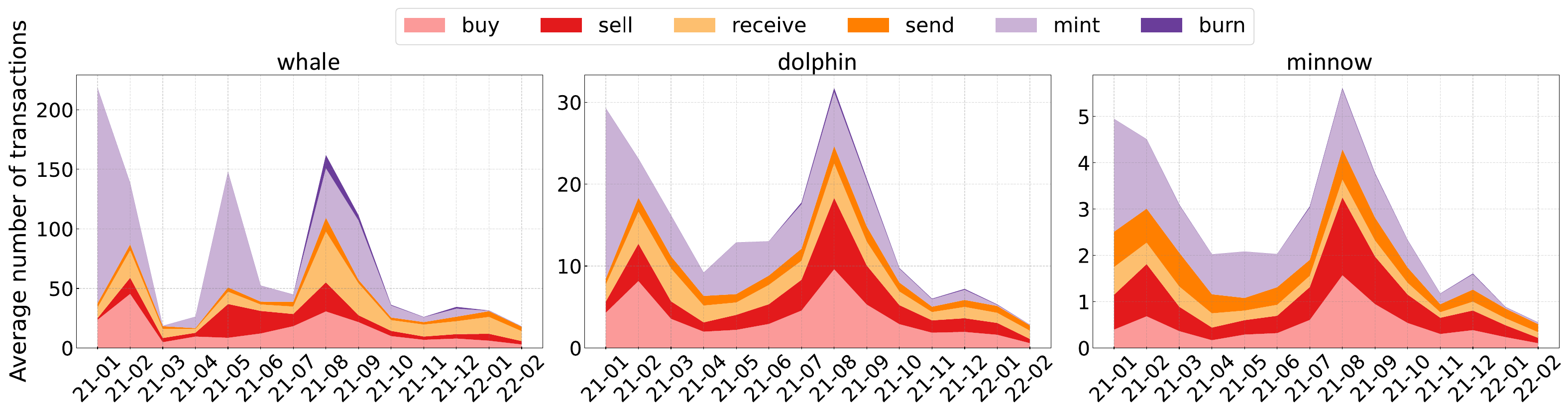}
  \caption{Monthly statistics of each transaction type of each group on average. (stacked plot)}
  \label{f:long_tx}
\end{figure*}

\subsection{Whales' Portfolio Holdings}
Common participants in the NFT market pursue financial gains. 
In this regard, examining their portfolio from various angles is highly valuable for measuring significant financial factors on the NFT ecosystem.
In this section, we examine the whales' NFT portfolio, preferences, and dominance in NFT collections.


\noindent \textbf{Top collection holdings.} We begin by searching for top 10 collections that whales mostly hold. Interestingly, we discover that the majority of the collections in the top 10 are popular collections with large trading volume.
The whales' most held collection is \textit{Art Blocks} (9.6K tokens held). Interestingly, whales hold 1.6K \textit{CryptoPunks} tokens which accounts for 16\% of total tokens in this collection. The number of tokens held by Whales in other collections can be found on Table \ref{t:whale_collections} of the Appendix.

\noindent\textbf{Price distribution.} To look into the price of each group's holdings, we divide the token price into six price points, ranging from \$0 to the maximum price in our data, which is \$23.3M.
Figure \ref{f:price_dist} describes the price distribution of tokens that each group held on February 28, 2022. Figure \ref{f:price_dist}(a) shows the distribution of tokens over total 91 collections. In addition, to look at distribution within a collection, \ref{f:price_dist}(b) shows the holding range of the most popular NFT project, \textit{CryptoPunks}. Note that the price distribution of popular collections (e.g., \textit{BAYC, Art Blocks}) also resembles the distribution of \textit{CryptoPunks}. Interestingly, almost all of the tokens with prices ranging over \$10M are in hands of whales. Whales also hold most of the tokens in the price range of \$1M to \$10M across all collections. On the whole, Figure \ref{f:price_dist} indicates that whales are the only holders of the high-price tokens, both for overall collections and the top collections.\\
\noindent\textbf{Findings and Insight.} 
Generally speaking, price of goods in a market is an indicator of its `value'. The result obtained in this section suggests that whales gain dominance over almost all highly valuable tokens.

\section{Whales' Impact on the NFT Market}
\label{s:impact}
To examine how influential whales are in the NFT ecosystem, we analyze the whales' trading behavior in terms of their impact on market sentiment and traders.

In Figure \ref{f:long_tx}, we observe that the average number of transactions by whales is overwhelmingly large. 
We observe larger fluctuations in the trading activities of whales than dolphins and minnows.
To understand the whales' behavior, we study how they respond to real-world events related to NFT.

\noindent{\bf Impact on market volatility during liquidation.}
Although the number of whales makes up a small portion of all traders, whales impact market volatility when liquidating NFT collections. Liquidation by whales has drastically increased market volatility. We discuss important events (E1$\sim$E3) in which the whales had a huge impact on the market.

First, we observe a rapid but short-lived peak in February 2021 in all groups. With the growing popularity of one collection, \textit{Hashmasks}~\cite{hashmasks}, traders buy a huge number of tokens from the collection, which accounts for more than half of their purchases. 

\textbf{(E1)} However, in April, NFT market conditions cooled abruptly. Traders in all groups barely participated in transactions due to a sharp decline (nearly 70\%) in the average price of NFTs~\cite{NFT_burst}. In fact, before the market crash, we observe whale activities drop to its lowest position in March. This implies that the decrease in whale activities has had a big impact on the NFT market sentiment.

\textbf{(E2)} In May, we observe an uplift in the number of sells only in the transaction graph for whales. In particular, the whales sell tokens that they have minted on this month. Interestingly, 95\% of such tokens are from two new collections, \textit{Meebits} and \textit{Bored Ape Yacht Club (BAYC)}. 
\textit{Meebits} received the spotlight even before launch, since it is made by the creators behind \textit{CryptoPunks}. 
On the other hand, \textit{BAYC} suddenly gained popularity due to a large quantity of whales' mintings~\cite{bayc_boom}. This promoted the sales of \textit{BAYC} tokens, especially in the dolphins and minnows group. In fact, \textit{BAYC} accounts for 40\% and 54\% of dolphins' and minnows' sales, respectively. 

\textbf{(E3)} We observe a steady increase in overall sales volume in July 2021, which increases dramatically by August. This is related to a surge in price of \textit{CryptoPunks} caused by whales. One day in August, a whale bought over 100 \textit{CryptoPunks} NFTs worth more than \$6M in total. Since then, the whales' purchase of the collection continued. Such whale activity sparked public interest in the NFT market, resulting in the influx of new traders~\cite{CryptoPunk_resurge}.

Except for specific periods in which whales actively liquidate their assets, whales tend to transfer (specifically, the act of receiving) instead of selling assets on the market. They prefer to hold assets relatively longer compared to dolphins and minnows. This is discussed in more detail in Section~\ref{s:strategy-hold}.

\noindent{\bf Leading investment trends. }
Whales have actively participated in \textit{mint} as early stage investors. Also, they prefer to invest specific collections that they are interested in, while other groups participate in a variety of collections. Here, we uncover several remarkable events as follows.

In January 2021, whales mint about four times more than any other transaction types, and most of them are \textit{Hashmasks}. Their mintings surged again in May, due to \textit{Meebits} and \textit{BAYC} as mentioned in E2. 
In the later months including August, some collections (e.g.,\textit{Punks Comic}) embed governance rights on tokens (a.k.a. DAO tokens), that increased the demand for mint. 
This suggests that whales are interested in exercising governance rights on a collection. 

NFT market capitalization is increasing due to rise in the number of collections and traders. However, we see a drop in practical trader activities compared to the early months. In fact, only 20\% of minnows participates in trading in the last month (see Table \ref{t:active_users} in the Appendix).  
This implies that the NFT market is mainly operated by whales and dolphins. In addition, we cannot observe any noticeable increase in whale transactions after August, which suggests the lack of emergence of influential collections (e.g., \textit{CryptoPunks}). Despite the large number of collection launches, only a few major collections account for the majority of the NFT market capitalization~\cite{nftgo_marketoverview}. Therefore, holding major collection NFTs allows the whales to maintain their position as whales.\\
\noindent{\bf Findings and Insight.}
The NFT market is predominantly being driven by whales. Whales have distinct trading patterns compared to that of other groups; whales have a huge impact on the market and often alter market sentiment.

\section{Deep-dive into Whales' investing strategies}
\label{s:strategy}
In this section, we discuss several investment strategies that whales utilize to achieve high financial gains. We discover three strategies: (i) minting or buying expensive NFTs and holding them for a long time until liquidation, (ii) intensively investing during the minting period, and (iii) taking advantage of self-trading.



\subsection{Long-term Investment}
\label{s:strategy-hold}
In a market, the value of a token is decided by its price; valuable tokens are limited but takes up a major portion of the market cap. Therefore, we define and track \textit{most-valuable NFTs} as the top 1\% most expensive NFTs from each collection. By accumulating such tokens, we obtain 14,192 most-valuable NFTs. 

\begin{figure}[t!]
  \centering
  \includegraphics[width=0.8\linewidth]{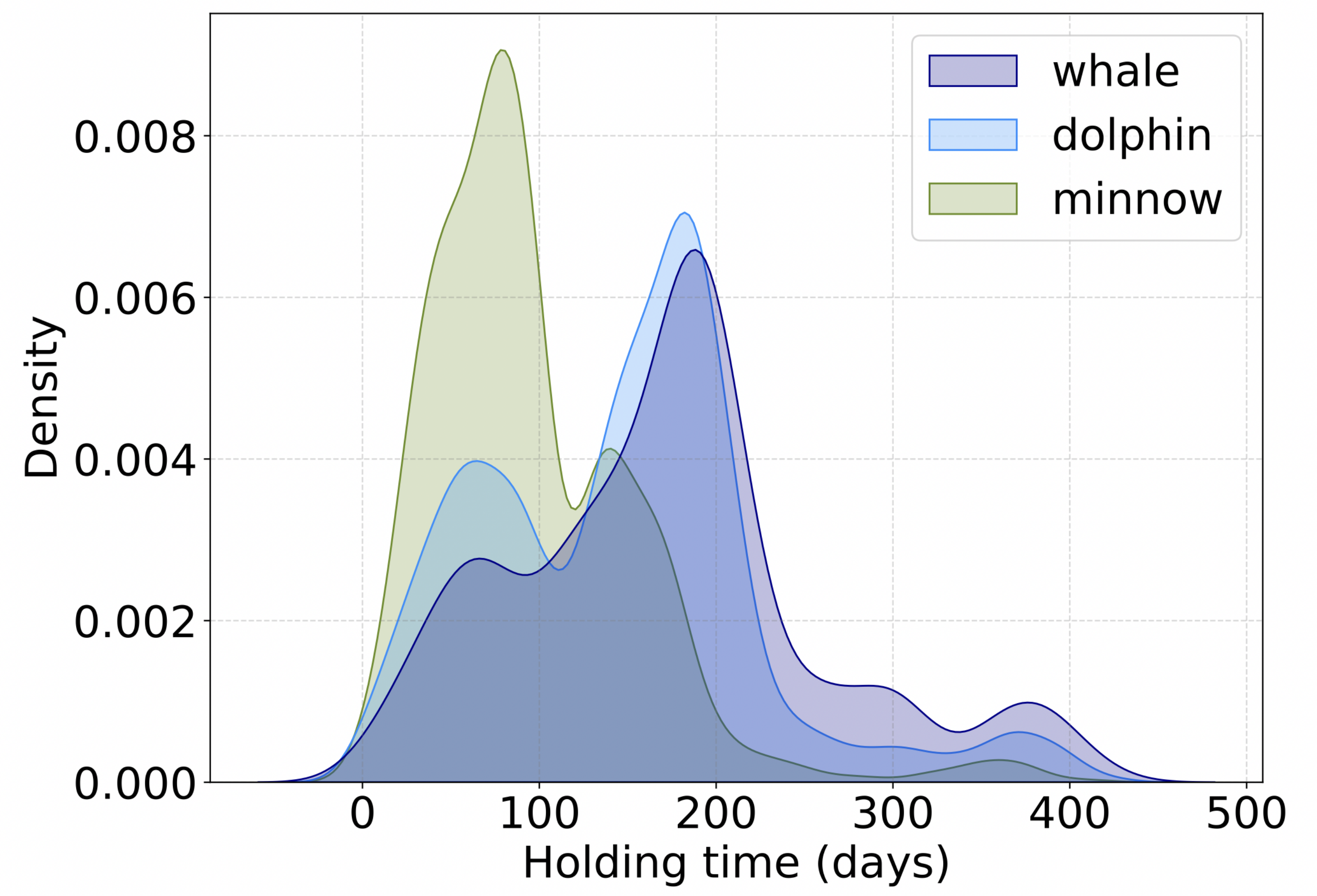}
  \caption{The holding time of most-valuable tokens per group}
  \label{f:holding_time}
\end{figure}

To investigate the strategies whales use on the most-valuable NFTs, we begin by calculating how long each group holds the tokens before selling them. Figure \ref{f:holding_time} shows the holding time of each group on most-valuable NFTs. Here, holding time is the duration between the last buy/receive time of each token and the last day of our data (February 28, 2022). As clearly shown from the graph, whales are likely to hold tokens longer than any other group. It turns out that some whales even hold tokens for the entire period of our data. On the other hand, for minnows, the peak in the left range of the graph is due to their relatively short holding period on collections that have low average token price. While the holding time distributions of minnows and whales are very distinct from each other, the distribution of dolphins resemble that of whales. Hence, the graph indicates that holding value leaders tend to hold the valuable tokens longer. 

\begin{figure}[t!]
  \centering
  \includegraphics[width=0.9\linewidth]{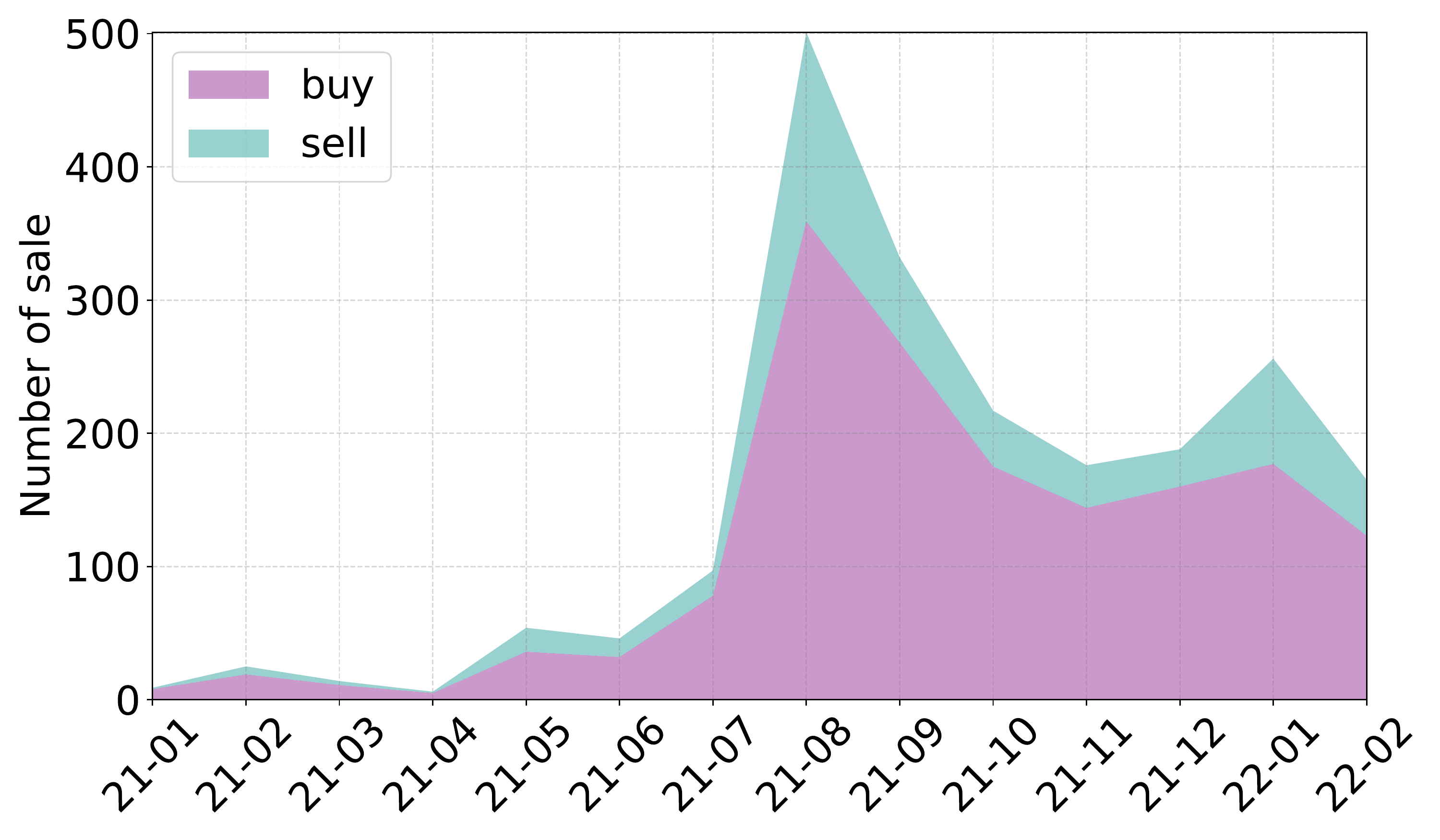}
  \caption{Buy/sell plot by whales on most-valuable tokens (stacked plot)}
  \label{f:mvt_sale}
  \vspace{-10pt}
\end{figure}

These observations bring about a question related to the liquidity of high-value tokens; do the other traders have any chance at acquiring such tokens? To answer this, we track the sale history of these tokens. Figure \ref{f:mvt_sale} illustrates the number of sales by whales on most-valuable NFTs. The graph clearly shows that the number of buys across all months heavily outweigh the number of sells. This indicates that high-value NFTs have low-liquidity. 

These two strategies are likely to be long-term tactics by whales to wait for valuable tokens to reach their desired price. Indeed, the results suggest that whales consistently collect most-valuable NFTs but barely sell them for a long duration. \\
\noindent\textbf{Findings and Insight.}
Whales are patient; they invest by holding valuable NFTs for a long period and do not sell them to other traders. NFT traders must be aware of low-liquidity of high-value tokens caused by whales.

\subsection{Deliberate Investment During the Minting Phase}
\begin{figure}[t!]
  \centering
  \includegraphics[width=0.6\linewidth]{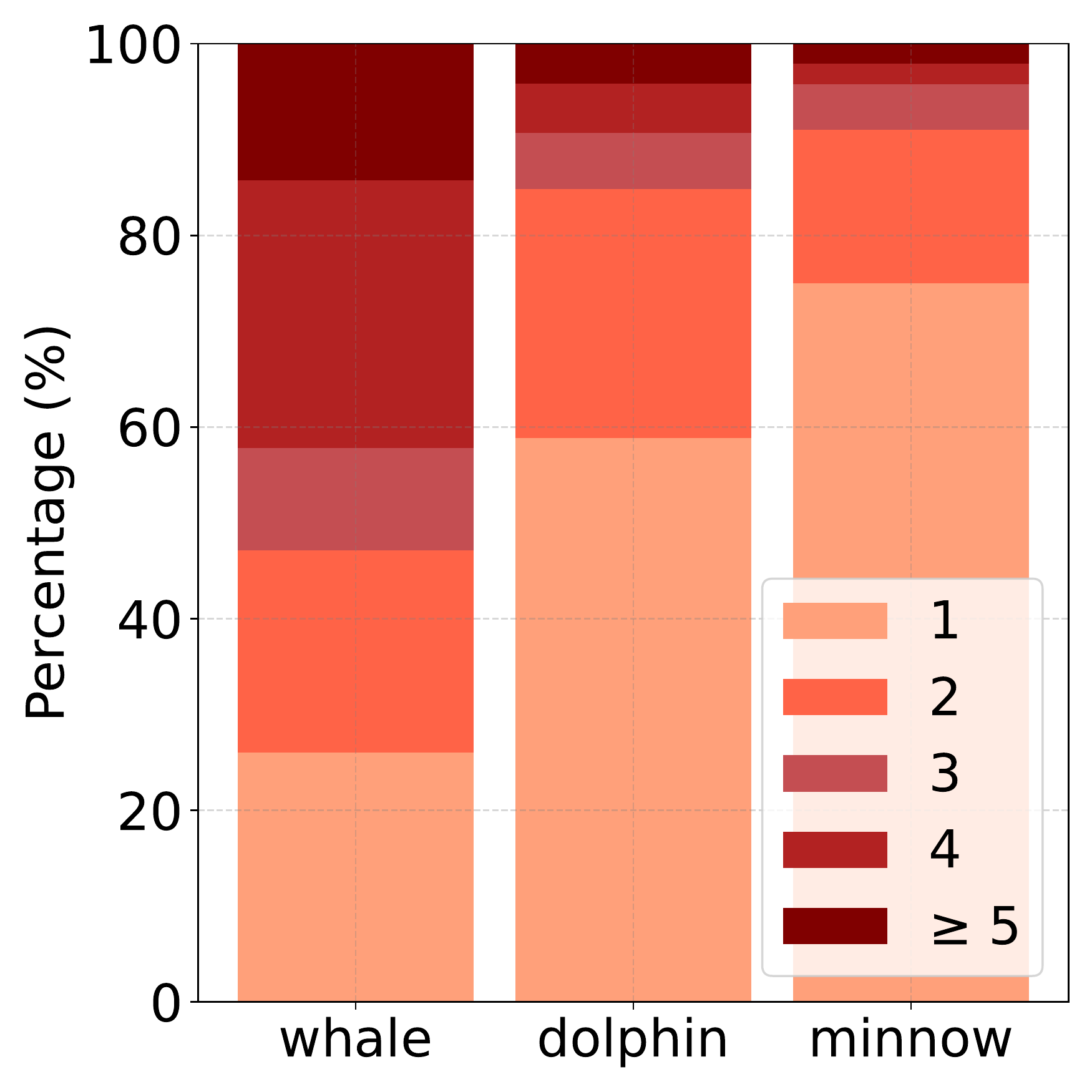}
  \caption{Number of transfers between \textit{mint} and first \textit{sell}}
  \label{f:mint_trasfer}
\end{figure}

We find out that whales take advantage of minting via unique methods. They use a number of strategies to raise the price of tokens they minted. We describe each strategy throughout this section.\\  
\noindent\textbf{Strategy 2-A.} The first strategy is related to the number of transfers between the first minting period and the first sale. Around 10\% of tokens minted by each group are transferred at least once before the first sale. To compare each group's transfer tendency, we look into those tokens in detail. Figure \ref{f:mint_trasfer} shows the number of transfers between the mint and the first sale, where the number is shown in percentage. As shown in the figure, a large portion of NFTs minted by whales are transferred many times (up to 25 times). 72.0\% of tokens minted by whales are transferred at least two times whereas the percentages are much smaller in dolphins and minnows. Furthermore, 13.8\% of tokens minted by whales go through transfer at least 5 times. 

\begin{figure}[t!]
  \centering
  \includegraphics[width=0.6\linewidth]{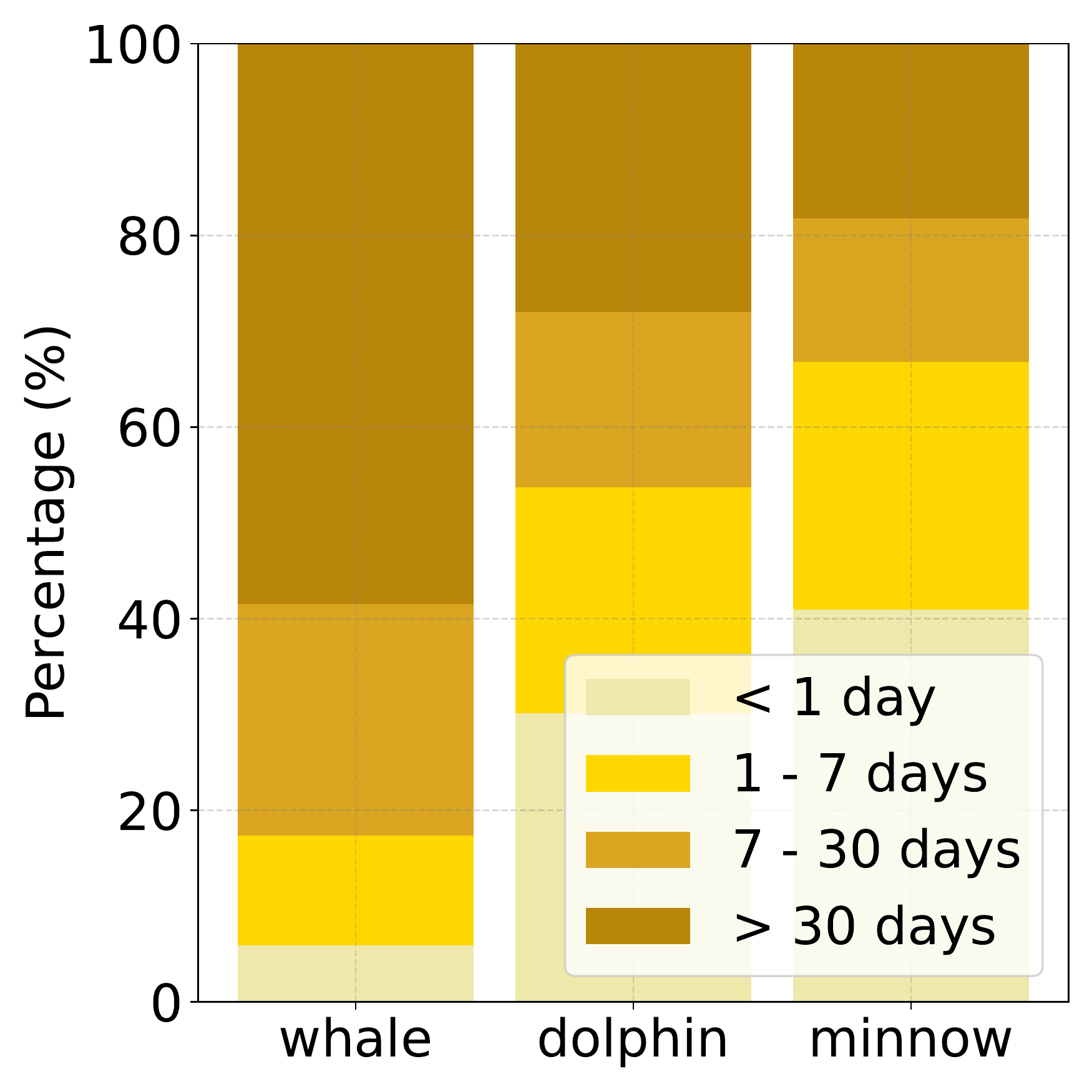}
  \caption{Time duration between \textit{mint} and first \textit{sell}}
  \label{f:mint_days}
  \vspace{-15pt}
\end{figure}

\noindent\textbf{Strategy 2-B.} Another noticeable strategy is relevant to time duration before the first sale. This can be seen from Figure \ref{f:mint_days} where the time duration is divided into four ranges: \textit{within a day}, \textit{a day to a week}, \textit{a week to a month} and \textit{more than a month}. Surprisingly, whales tend to wait for long periods of time before the first sale; 82.6\% of tokens take more than a week and 58.5\% take more than a month. In contrast, dolphins and minnows wait less and the majority of tokens are sold within a week for both groups.

\noindent\textbf{Strategy 2-C.} This strategy is used by some of the minters. Usual sales after minting do not involve the address (i.e., minter) that minted the token. However, we find that some minters later receive back the token that they minted. To give an instance that occurred in September 2021, the minter first minted the token and sold it to another wallet address for 6.25 ETH. Then, the token was transferred back to the minter. Finally, the minter sold the token to another trader for 6 ETH. Indeed, we find that 13.7K tokens are involved in this type of transaction pattern. Whales and dolphins are involved in 10.4K tokens which accounts for 77.4\% of such tokens. This is a significant number considering the fact that usually the total number of tokens for a collection is around 10K. 

All three strategies are effectively utilized to maximize the profit of whales. We find out that the profit of the first sale is proportional to all of these strategies. Details of these profits can be found in Table \ref{t:mint_appendix1} and \ref{t:mint_appendix2}. Further details for mint profits from whales are discussed in Section \ref{s:profit-mint}\\
\noindent\textbf{Findings and Insight.} Whales use their own special strategies to maximize the profit of tokens they minted. 


\subsection{Wash Trading}
\begin{figure}[t!]
  \centering
  \includegraphics[width=0.9\linewidth]{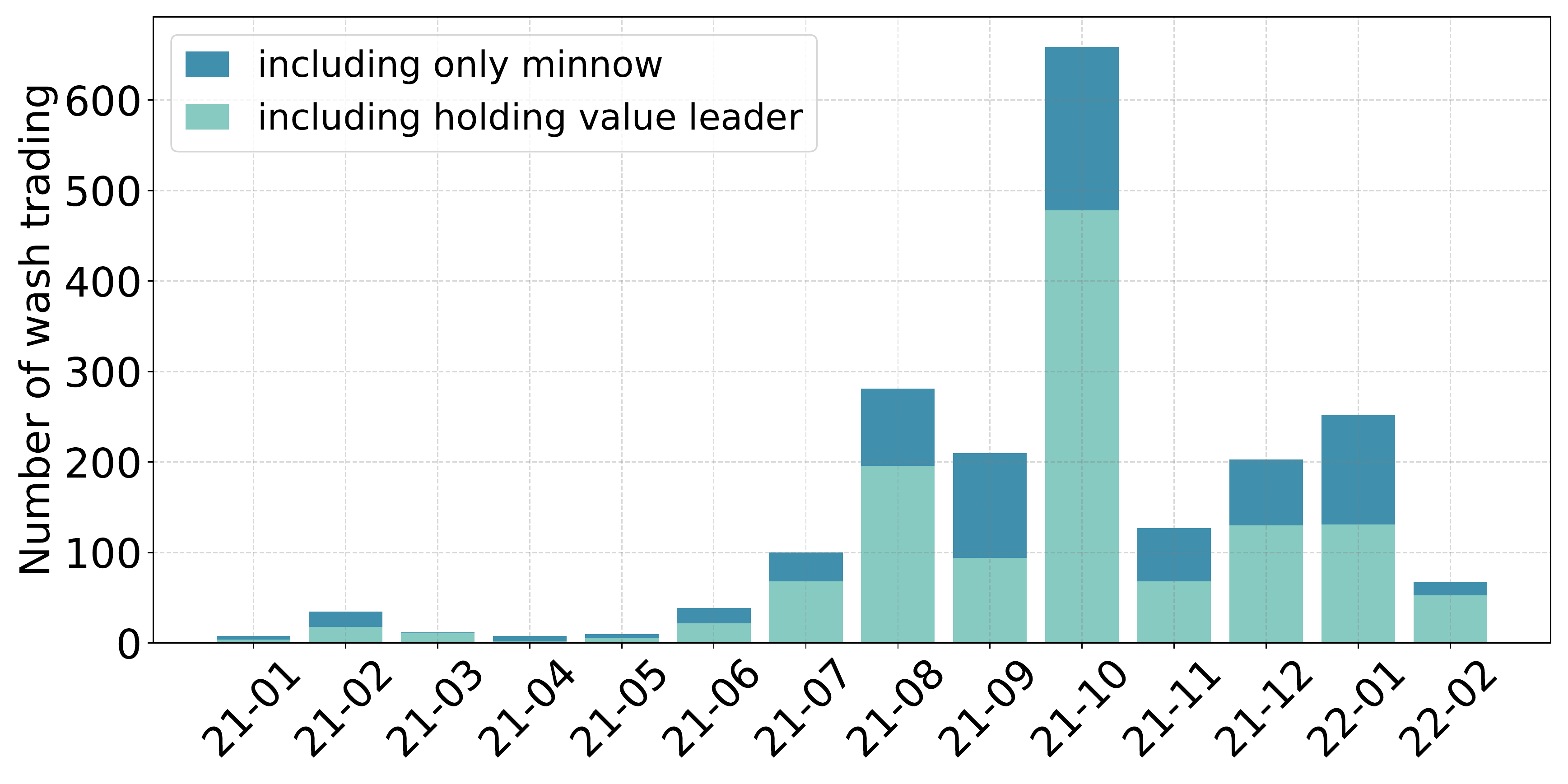}
  \caption{Number of wash trading instances(stacked plot)}
  \label{f:long_wash}
  \vspace{-15pt}
\end{figure}
Wash trading is a collusion by the buyer and the seller to artificially inflate the trading volume of an asset. They repeatedly trade their assets between them, which results in cycles in the sale graph, where nodes are traders and edges are sales between them. Therefore, wash tradings are captured by finding strongly connected components (SCCs) in each token sale graph~\cite{victor2021detecting}. 
Indeed, multiple unconditional token transfers can be used in trading malpractices to avoid monitoring~\cite{das2021understanding, victor2021detecting}. 
Thus, we construct each NFT transaction graph, where nodes are traders and edges are transactions between them, and find SCCs in each graph including sales at least once. 

We detect 3,558 instances of wash trading in 1,676 NFTs across 82 out of 91 collections including 3,676 traders in our data. Note that the collected NFT transaction data is different for each period and for each collection, the number of wash trading instances can be different from the work done by Victor et al.~\cite{victor2021detecting}. Surprisingly, the holding value leaders perform 69\% of wash trading. Furthermore, 42 whales, which makes up 10\% of all whales, are involved. We find that popular collections with high trading volume (e.g., \textit{CryptoPunks, BAYC}) also belong to manipulations done by the holding value leaders.

To understand when and why wash trading occurs over time, we locate SCCs in each token transaction graph from the monthly transaction records.  
Figure \ref{f:long_wash} shows the number of wash trading over time.
First, we observe a noticeable peak in October. Wash trading for \textit{The n project} started in September (when it was launched) and recorded 506 cases in October. We confirm that the average price of the collection tokens traded sharply rises whenever washing trading is involved as shown in Figure \ref{fig:wash_price} of the Appendix. 

This is a common phenomenon in newly launched collections. In order for a newly-launched collection to be verified by \textit{Opensea} (the largest NFT marketplace), the trading volume must be over 100 ETH~\cite{opensea_verify}. For this reason, many traders are tempted to perform wash trading. Similarly, we can see a increase in the number of wash trading in August 2021, when many new collections emerged. 

The number of wash trading shows another small peak in January 2022, which is related to the policy of LooksRare~\cite{meebits_wash}, a newly-launched NFT marketplace. It rewards valuable tokens to traders according to each trader’s trading volume, which lures traders into wash trading. \\
\noindent\textbf{Findings and Insight.}
Whales and dolphins actively participate in wash trading to receive verification of newly-launched collections in NFT matketplaces or earn rewards by abusing NFT market policy. NFT prices can be driven up and down along with their wash trading. Popular collections (e.g., \textit{CryptoPunks, BAYC}) are also unavoidable.

\section{Evaluating Investment Performance}
\label{s:profit}


In this section, we evaluate the investment performance of each group and discuss how whales achieve high returns compared to other market participants.

\subsection{Investment Performance}
\begin{figure}[t!]
  \centering
  \includegraphics[width=0.8\linewidth]{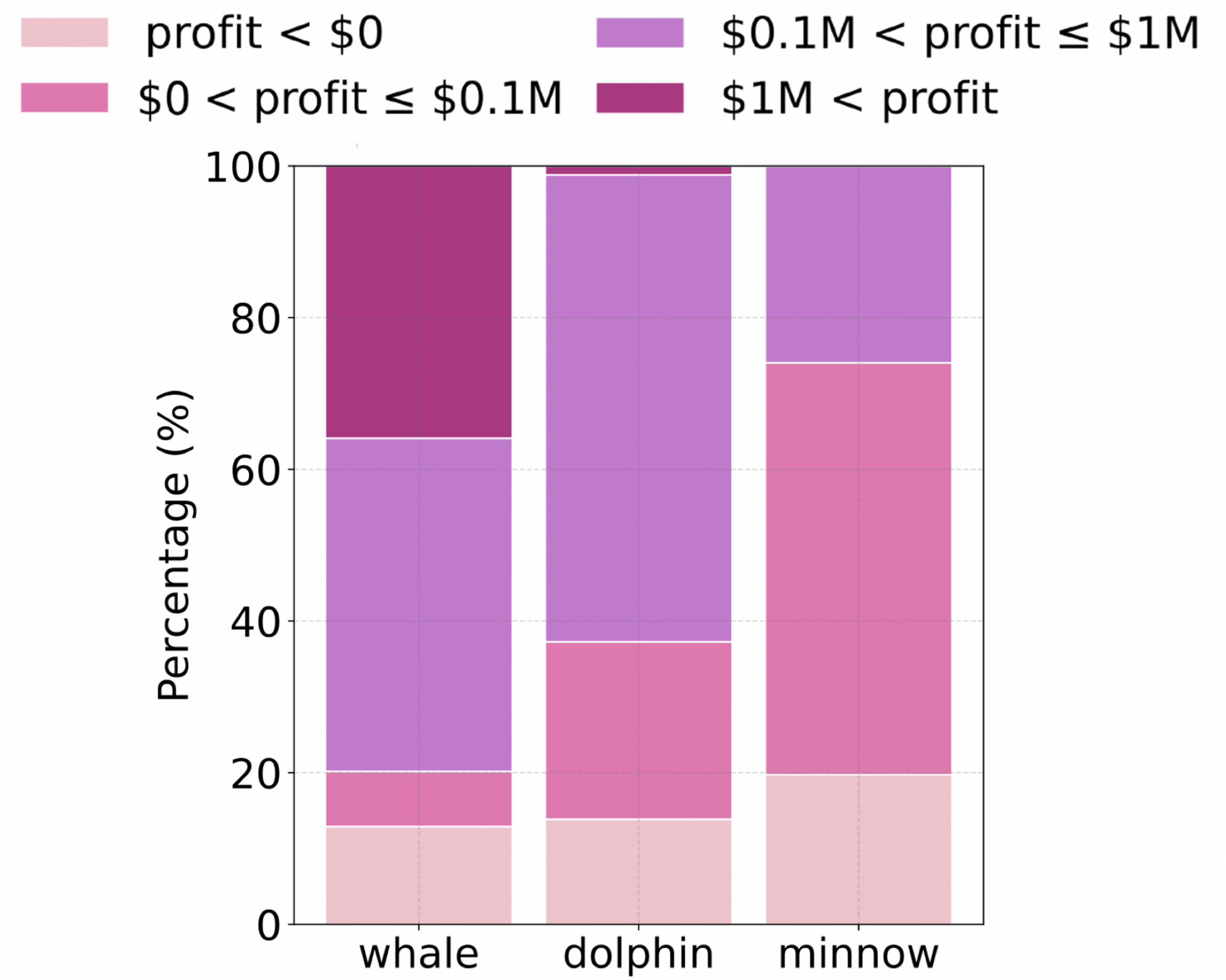}
  \caption{Investment performance per trading group}
  \label{f:profit_basic}
  \vspace{-10pt}
\end{figure}

Figure \ref{f:profit_basic} illustrates the investment performance of each trading group. We divide the profit range with a log scale due to the wide range of profit among traders. Note that we do not include traders who have never participated in profit activity (i.e., zero profit). The trader who gained the maximum profit is a whale (\$18.9M) and the trader with the largest loss is a dolphin (-\$2.7M). The most noticeable observation is that 35.9\% of traders who belong to whales make profits larger than \$1M while there hardly exists any from dolphins and minnows. Moreover, while 79.8\% of whales produced profits greater than \$100K, the percentages are much lower in other groups. This suggests that a large fraction of whales produces a considerably higher profit compared to any other group.\\
\noindent\textbf{Findings and Insight.}
We identify that whales have achieved significant financial gain. A large fraction of whales achieved profit over \$1M which shows great contrast with other groups.

\subsection{Sources of Financial Gains from Whales}
Finally, we scrutinize the source of profit from whales. We divide the profit source into two parts: buy profit and mint profit. 




\noindent\textbf{Buy profit.} This type of profit occurs from ordinary sales, except the very first sale of a token. For profit generated from buy, we focus on the collections that whales take advantage of. Whales gained profit from 74 collections out of 91 in total. Among them, we sort the top collections in which the whales were most profitable. If a collection does not generate any profit for several months (e.g., launching in the middle of the entire period), we only mark zero profit once for graph's visibility. 

\begin{figure*}[t!]
  \centering
  \includegraphics[width=0.9\linewidth]{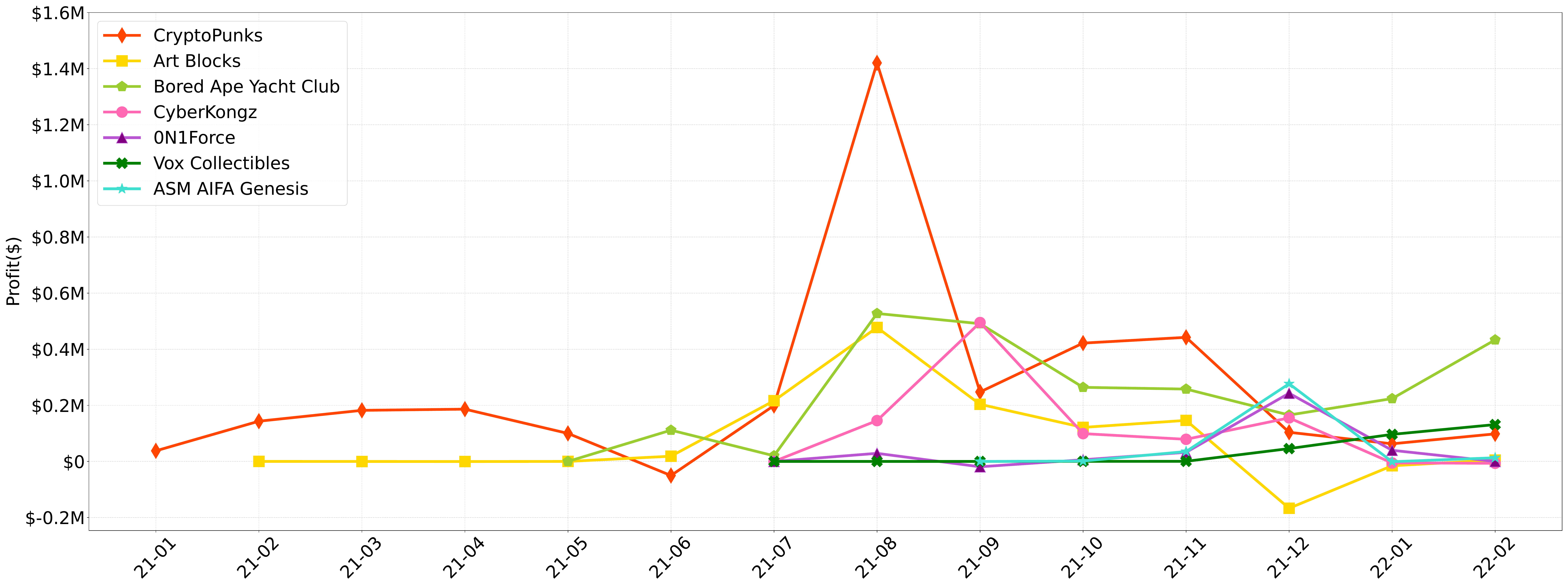}
  \caption{Average Profit of whales per collection}
  \label{f:col_profit}
  \vspace{-10pt}
\end{figure*}

Figure \ref{f:col_profit} shows the average profit of whales from the top 7 collections (top 10\% of 74 collections) for each month. Overall, the majority of these collections are well-known for their large trading volume. Among them, profits from \textit{CryptoPunks, Art Blocks} and \textit{BAYC} are consistently large, which indicates that whales continuously gain profit from very popular collections. For the peak of the rest of the collections, the profit ranks the top due to the selling of several expensive tokens by small number of whales. For example, a peak of \textit{ASM AIFA Genesis} on December 2021 was generated by a single whale that sold 56 tokens.

\noindent\textbf{Mint profit.}
Whales consistently gained profit from mint profit, which is the profit from the first sale just after mint. Surprisingly, although mint profit occurs only once per token, mint profits are comparable to buy profits and even are larger on some months. This is noteworthy as, 1) the number of mint logs accounts for 28.2\% of total transfer logs in our data (1.1M out of 3.9M) and 2) most collections have a limited number of tokens available for mint (e.g., only 10K tokens for \textit{BAYC}).  

Until July 2021, whales sell the tokens that they have minted from a limited number of collections. For example, The mint profit of whales consistently occur from \textit{Hashmasks} in the first half of the year. However, starting from August where the number of traders in the NFT market drastically increased, whales start to gain mint profit from a wide range of collections. This is likely to be due to the launch of many collections after August. Nevertheless, most of the mint profit are obtained from popular collections mentioned above, e.g., \textit{Art Blocks, Meebits, BAYC} until the last month (February 2022).

Meanwhile, some collections utilize \textit{airdrop} as a way to advertise themselves during the launch period; they let traders mint the tokens without paying any mint fees \cite{airdrop}. To amplify the effects of advertising, some collections even choose to send their tokens to popular wallet addresses in a unilateral way. Since whale accounts are easily available in NFT platforms (e.g.,NFTGo \cite{nftgo}), it is likely for whales to receive lots of tokens through airdrop. Therefore, we investigate the profits produced by whales through airdrop. 

\begin{figure}[h!]
  \centering
  \includegraphics[width=0.9\linewidth]{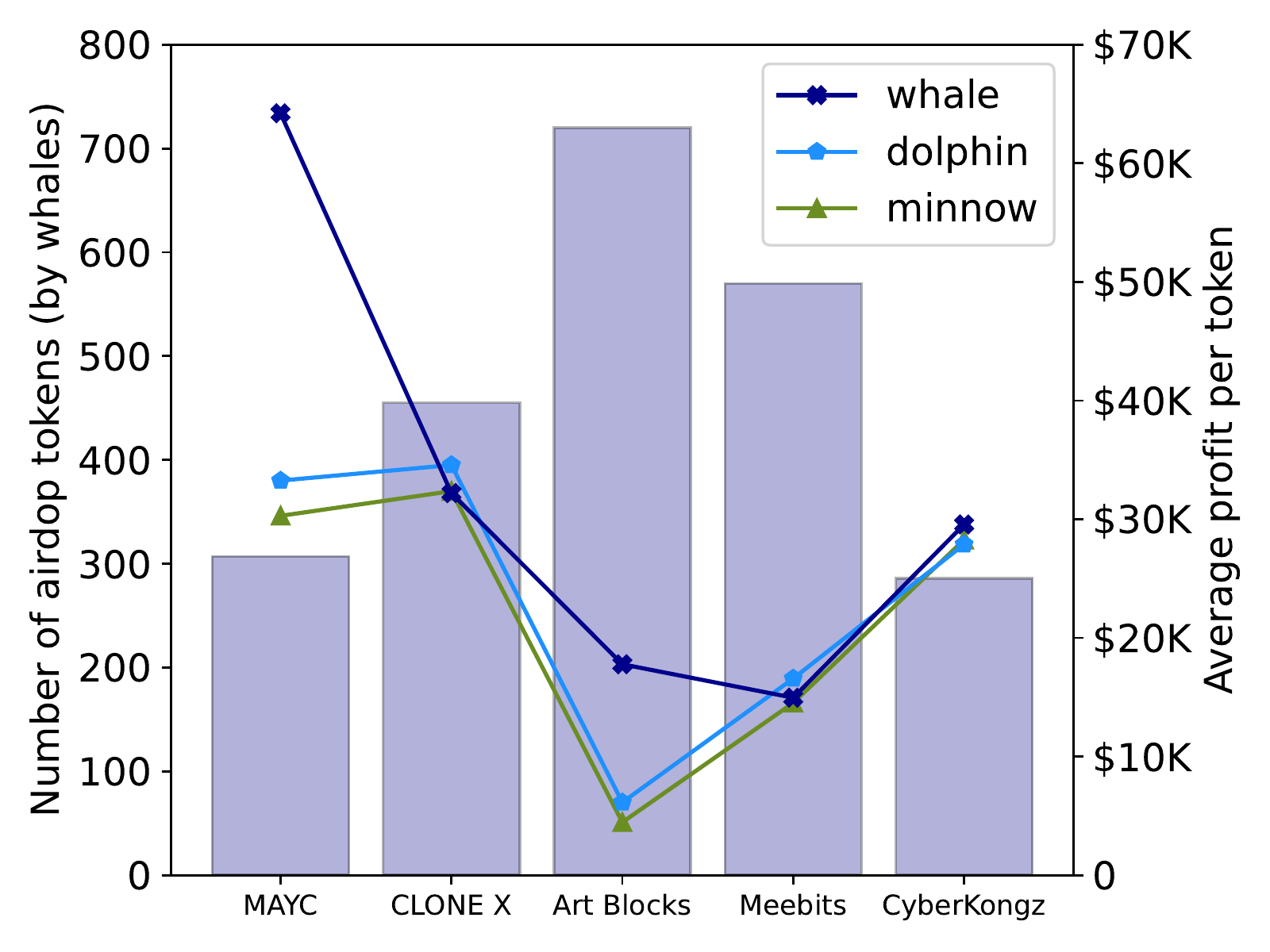}
  \caption{Average Profit of airdrop tokens (\textit{MAYC} and \textit{CLONE X} are abbreviations for \textit{Mutant Ape Yacht Club} and \textit{RTFKT CLONE X + Murakami}, respectively.)}
  \label{f:airdrop}
  \vspace{-10pt}
\end{figure}

Figure \ref{f:airdrop} shows the top 5 collections in which the total profit of whales is the largest. The number of airdrop tokens sold by whales take up 9 to 20\% of airdrop tokens in these collections. This is a large portion considering the small number of whales and implies that whales actively participate in selling airdrop tokens. Interestingly, the five collections in Figure \ref{f:airdrop} are very popular collections with large trading volumes. Overall, among the 41 collections from which the whales obtained profit through airdrop, we find that the tokens sold by whales are generally more expensive than tokens by other groups on 36 collections. This indicates two possibilities: 1) whales already have the public's trust in the NFT market and the traders are willing to pay high prices for the tokens minted by whales or 2) whales receive relatively high-value tokens of a collection via airdrop. Still, in either case, it is an undeniable fact that whales have power in the minting process that can lead to larger profit while no others can. \\
\noindent\textbf{Findings and Insight.} Whales are mainly lucrative through selling NFTs of popular collections. More importantly, they actively participate in minting process, sell NFTs at higher price than any other group. This allows whales to become successful investors. 
\label{s:profit-mint}

\section{Discussion}
\label{s:discussion}

Throughout this paper, we find out that whales are highly influential traders in the NFT ecosystem. Whales hold a mere 5 percent of all NFT tokens on the Ethereum blockchain, but their worth accounts for nearly 20 percent of the entire NFT market value. In addition, although the NFT market features relatively wide variations in prices, nearly all of the high value items worth more than one million dollars belong to the whales. This means that whales exclusively own almost all of the dominant NFT items. Considering that a small number of NFT projects with a high market cap (e.g., \textit{CryptoPunks}, \textit{BAYC}, \textit{Doodles}) is bringing success to the NFT ecosystem, to say that the NFT market is being driven by whales is not an overstatement.

In summary, the average profit from whales is far greater than that of the other groups. In addition, the most highly profitable traders (i.e., make profits of more than \$1M) make up over 35\% of all whale traders, while there such traders only make up under one percent of dolphins and minnows. A large portion of traders from minnows did not make profits, with some even faced with considerable losses. This indicates that a small number of whales (note that we only consider the top 0.1 percent of traders as whales) takes most of the rewards for success in the NFT market. In other words, the NFT ecosystem is a harsh environment to find success for the majority of the traders involved (non-whales), and stakes are high for the minority traders. Therefore, it is imperative that various studies about the NFT ecosystem are conducted and measures to protect investors are established.

\section{Related Work}
\label{s:rw}
To the best our knowledge, this is the first longitudinal study of NFT ecosystem that focuses on identifying and characterizing market
movers from a financial point of view.
Wang \textit{et al.}~\cite{wang2021non} provides an overview of the NFT ecosystem with regard to the technical components, etc. Our work is different in that we analyze how NFT market operates from a financial standpoint.
Dowling \textit{et al.}~\cite{dowling2022non} and Ante \textit{et al.}~\cite{ante2021non} study correlation between price of NFTs and cryptocurrencies, and shows that they have a low correlation. 
Distinct from their works, we focus on the NFT market driven by traders. 
Brunet \textit{et al.}~\cite{casale2021networks} and Nandi \textit{et al.}~\cite{nadini2021mapping} study the basic topological structure of NFT trading networks and show that each is similar to other social networks. 
Das \textit{et al.}~\cite{das2021understanding} presents a systematic overview of how the NFT ecosystem works and uncovers potential security issues. 
Wachter \textit{et al.}~\cite{von2022nft} also examines potentially illicit trading patterns in the NFT market with two detection algorithms and their effect on price. Still, there is no understanding on how malicious behaviors have changed over time and analysis on the performers of wash trading.

\section{Conclusion}
\label{s:conclusion}
Following the rapid growth of cryptocurrencies, NFTs have recently received wide attention from the public. Many participants have successfully traded NFT items with financial gain, leading to the emergence of NFT whales. However, many participants suffer from the lack of a clear viewpoint on the NFT ecosystem.
In this research, we perform the first longitudinal study to construct in-depth analysis on the NFT market, focusing on unique activities of market participants including whales and market movements. Specifically, we reveal whales on the NFT ecosystem and discover that they possess unique strategies to maximize their earnings from NFT trades and greatly impact the entire market. 
Consequently, we believe that our findings and insight in this work shed light on the NFT ecosystem, which has been minimally investigated to date.


\bibliographystyle{ACM-Reference-Format}
\bibliography{reference}


\begin{thebibliography}{27}


\ifx \showCODEN    \undefined \def \showCODEN     #1{\unskip}     \fi
\ifx \showDOI      \undefined \def \showDOI       #1{#1}\fi
\ifx \showISBNx    \undefined \def \showISBNx     #1{\unskip}     \fi
\ifx \showISBNxiii \undefined \def \showISBNxiii  #1{\unskip}     \fi
\ifx \showISSN     \undefined \def \showISSN      #1{\unskip}     \fi
\ifx \showLCCN     \undefined \def \showLCCN      #1{\unskip}     \fi
\ifx \shownote     \undefined \def \shownote      #1{#1}          \fi
\ifx \showarticletitle \undefined \def \showarticletitle #1{#1}   \fi
\ifx \showURL      \undefined \def \showURL       {\relax}        \fi
\providecommand\bibfield[2]{#2}
\providecommand\bibinfo[2]{#2}
\providecommand\natexlab[1]{#1}
\providecommand\showeprint[2][]{arXiv:#2}

\bibitem[loo(2022)]%
        {looksrare}
 \bibinfo{year}{2022}\natexlab{}.
\newblock \bibinfo{title}{Looksrare}.
\newblock \bibinfo{howpublished}{\url{https://looksrare.org/}}.
\newblock
\newblock
\shownote{Accessed: 2022-05-19}.


\bibitem[ope(2022a)]%
        {opensea}
 \bibinfo{year}{2022}\natexlab{a}.
\newblock \bibinfo{title}{Opensea}.
\newblock \bibinfo{howpublished}{\url{https://opensea.io/}}.
\newblock
\newblock
\shownote{Accessed: 2022-05-19}.


\bibitem[ope(2022b)]%
        {opensea_verify}
 \bibinfo{year}{2022}\natexlab{b}.
\newblock \bibinfo{title}{What is a verified account or badged collection?}
\newblock
  \bibinfo{howpublished}{\url{https://support.opensea.io/hc/en-us/articles/360063519133-What-is-a-verified-account-or-badged-collection-}}.
\newblock


\bibitem[Ante(2021)]%
        {ante2021non}
\bibfield{author}{\bibinfo{person}{Lennart Ante}.}
  \bibinfo{year}{2021}\natexlab{}.
\newblock \showarticletitle{The non-fungible token (NFT) market and its
  relationship with Bitcoin and Ethereum}.
\newblock \bibinfo{journal}{\emph{Available at SSRN 3861106}}
  (\bibinfo{year}{2021}).
\newblock


\bibitem[Banton(2021)]%
        {bitcoin_whale}
\bibfield{author}{\bibinfo{person}{Caroline Banton}.}
  \bibinfo{year}{2021}\natexlab{}.
\newblock \bibinfo{title}{Bitcoin Whale}.
\newblock
  \bibinfo{howpublished}{\url{https://www.investopedia.com/terms/b/bitcoin-whale.asp}}.
\newblock


\bibitem[Browne(2022)]%
        {nft_boom2}
\bibfield{author}{\bibinfo{person}{Ryan Browne}.}
  \bibinfo{year}{2022}\natexlab{}.
\newblock \bibinfo{title}{Trading in NFTs spiked 21,000\% to more than \$17
  billion in 2021, report says}.
\newblock
  \bibinfo{howpublished}{\url{https://www.cnbc.com/2022/03/10/trading-in-nfts-spiked-21000percent-to-top-17-billion-in-2021-report.html}}.
\newblock


\bibitem[Casale-Brunet et~al\mbox{.}(2021)]%
        {casale2021networks}
\bibfield{author}{\bibinfo{person}{S Casale-Brunet}, \bibinfo{person}{P
  Ribeca}, \bibinfo{person}{P Doyle}, {and} \bibinfo{person}{M Mattavelli}.}
  \bibinfo{year}{2021}\natexlab{}.
\newblock \showarticletitle{Networks of Ethereum Non-Fungible Tokens: A
  graph-based analysis of the ERC-721 ecosystem}. In
  \bibinfo{booktitle}{\emph{2021 IEEE International Conference on Blockchain
  (Blockchain)}}. IEEE, \bibinfo{pages}{188--195}.
\newblock


\bibitem[Das et~al\mbox{.}(2021)]%
        {das2021understanding}
\bibfield{author}{\bibinfo{person}{Dipanjan Das}, \bibinfo{person}{Priyanka
  Bose}, \bibinfo{person}{Nicola Ruaro}, \bibinfo{person}{Christopher Kruegel},
  {and} \bibinfo{person}{Giovanni Vigna}.} \bibinfo{year}{2021}\natexlab{}.
\newblock \showarticletitle{Understanding Security Issues in the NFT
  Ecosystem}.
\newblock \bibinfo{journal}{\emph{arXiv preprint arXiv:2111.08893}}
  (\bibinfo{year}{2021}).
\newblock


\bibitem[Dowling(2022)]%
        {dowling2022non}
\bibfield{author}{\bibinfo{person}{Michael Dowling}.}
  \bibinfo{year}{2022}\natexlab{}.
\newblock \showarticletitle{Is non-fungible token pricing driven by
  cryptocurrencies?}
\newblock \bibinfo{journal}{\emph{Finance Research Letters}}
  \bibinfo{volume}{44} (\bibinfo{year}{2022}), \bibinfo{pages}{102097}.
\newblock


\bibitem[Google(2022)]%
        {hashmasks}
\bibfield{author}{\bibinfo{person}{Google}.} \bibinfo{year}{2022}\natexlab{}.
\newblock \bibinfo{title}{Google Search Trend - Hashmasks}.
\newblock
  \bibinfo{howpublished}{\url{https://trends.google.com/trends/explore?date=2021-01-01\%202022-02-28&q=Hashmasks}}.
\newblock


\bibitem[Graham(2022)]%
        {airdrop}
\bibfield{author}{\bibinfo{person}{Michelai Graham}.}
  \bibinfo{year}{2022}\natexlab{}.
\newblock \bibinfo{title}{Crypto Airdrops Explained}.
\newblock
  \bibinfo{howpublished}{\url{https://boardroom.tv/airdrop-crypto-nft-explained/}}.
\newblock


\bibitem[Gupta(2022)]%
        {bitcoin_whale2}
\bibfield{author}{\bibinfo{person}{Alinda Gupta}.}
  \bibinfo{year}{2022}\natexlab{}.
\newblock \bibinfo{title}{What Are Whales and How Do They Manipulate
  Cryptocurrency?}
\newblock
  \bibinfo{howpublished}{\url{https://www.jumpstartmag.com/what-are-whales-and-how-do-they-manipulate-cryptocurrency/}}.
\newblock


\bibitem[Hollander(2020)]%
        {event_logs}
\bibfield{author}{\bibinfo{person}{Luit Hollander}.}
  \bibinfo{year}{2020}\natexlab{}.
\newblock \bibinfo{title}{Understanding event logs on the Ethereum blockchain}.
\newblock
  \bibinfo{howpublished}{\url{https://medium.com/mycrypto/understanding-event-logs-on-the-ethereum-blockchain-f4ae7ba50378}}.
\newblock


\bibitem[Li(2021)]%
        {CryptoPunk_resurge}
\bibfield{author}{\bibinfo{person}{Joyce Li}.} \bibinfo{year}{2021}\natexlab{}.
\newblock \bibinfo{title}{CryptoPunk Prices See 53\% Increase After Ethereum
  Whales' Purchase Frenzy}.
\newblock
  \bibinfo{howpublished}{\url{https://hypebeast.com/2021/8/cryptopunk-price-rises-after-ethereum-whales-purchase-frenzy}}.
\newblock


\bibitem[Major(2021)]%
        {nft_august}
\bibfield{author}{\bibinfo{person}{Jordan Major}.}
  \bibinfo{year}{2021}\natexlab{}.
\newblock \bibinfo{title}{Global interest in NFTs soars by 426\% in a month,
  data shows}.
\newblock
  \bibinfo{howpublished}{\url{https://finbold.com/global-interest-in-nfts-soars-by-426-in-a-month-data-shows/}}.
\newblock


\bibitem[Mattei(2021)]%
        {nft_boom1}
\bibfield{author}{\bibinfo{person}{Shanti Escalante-De Mattei}.}
  \bibinfo{year}{2021}\natexlab{}.
\newblock \bibinfo{title}{The Year of the NFT: How an Emerging Medium Went
  Mainstream in 2021}.
\newblock
  \bibinfo{howpublished}{\url{https://www.artnews.com/list/art-news/artists/2021-year-of-the-nft-1234614022/}}.
\newblock


\bibitem[Nadini et~al\mbox{.}(2021)]%
        {nadini2021mapping}
\bibfield{author}{\bibinfo{person}{Matthieu Nadini}, \bibinfo{person}{Laura
  Alessandretti}, \bibinfo{person}{Flavio Di~Giacinto}, \bibinfo{person}{Mauro
  Martino}, \bibinfo{person}{Luca~Maria Aiello}, {and} \bibinfo{person}{Andrea
  Baronchelli}.} \bibinfo{year}{2021}\natexlab{}.
\newblock \showarticletitle{Mapping the NFT revolution: market trends, trade
  networks, and visual features}.
\newblock \bibinfo{journal}{\emph{Scientific reports}} \bibinfo{volume}{11},
  \bibinfo{number}{1} (\bibinfo{year}{2021}), \bibinfo{pages}{1--11}.
\newblock


\bibitem[NFTBank(2022)]%
        {nftbank}
\bibfield{author}{\bibinfo{person}{NFTBank}.} \bibinfo{year}{2022}\natexlab{}.
\newblock \bibinfo{title}{{NFTBank}}.
\newblock \bibinfo{howpublished}{\url{https://nftbank.ai/}}.
\newblock
\newblock
\shownote{Accessed: 2022-05-18}.


\bibitem[NFTGo(2022a)]%
        {nftgo_marketoverview}
\bibfield{author}{\bibinfo{person}{NFTGo}.} \bibinfo{year}{2022}\natexlab{a}.
\newblock \bibinfo{title}{Market Overview}.
\newblock
  \bibinfo{howpublished}{\url{https://nftgo.io/analytics/market-overview/}}.
\newblock
\newblock
\shownote{Accessed: 2022-05-18}.


\bibitem[NFTGo(2022b)]%
        {nftgo}
\bibfield{author}{\bibinfo{person}{NFTGo}.} \bibinfo{year}{2022}\natexlab{b}.
\newblock \bibinfo{title}{{NFTGo}}.
\newblock \bibinfo{howpublished}{\url{https://nftgo.io/}}.
\newblock
\newblock
\shownote{Accessed: 2022-05-18}.


\bibitem[Onanuga(2021)]%
        {bitcoin_whale3}
\bibfield{author}{\bibinfo{person}{Tola Onanuga}.}
  \bibinfo{year}{2021}\natexlab{}.
\newblock \bibinfo{title}{Bitcoin whales: what are they – and how are they
  affecting the cryptocurrency's price?}
\newblock
  \bibinfo{howpublished}{\url{https://www.businessinsider.com/bitcoin-whales-the-key-facts-figures-you-need-to-know-2021-1}}.
\newblock


\bibitem[Shaw(2021)]%
        {NFT_burst}
\bibfield{author}{\bibinfo{person}{Anny Shaw}.}
  \bibinfo{year}{2021}\natexlab{}.
\newblock \bibinfo{title}{Sorry to burst your bubble: NFT prices slump 70\%}.
\newblock
  \bibinfo{howpublished}{\url{https://www.theartnewspaper.com/2021/04/08/sorry-to-burst-your-bubble-nft-prices-slump-70percent}}.
\newblock


\bibitem[Trash(2021)]%
        {bayc_boom}
\bibfield{author}{\bibinfo{person}{Corporate Trash}.}
  \bibinfo{year}{2021}\natexlab{}.
\newblock \bibinfo{title}{What is Bored Ape Yacht Club? The Ape NFT
  Transforming NFTs}.
\newblock
  \bibinfo{howpublished}{\url{https://momentranks.com/blog/what-is-bored-ape-yacht-club-the-ape-nft-transforming-nfts}}.
\newblock


\bibitem[Victor and Weintraud(2021)]%
        {victor2021detecting}
\bibfield{author}{\bibinfo{person}{Friedhelm Victor} {and}
  \bibinfo{person}{Andrea~Marie Weintraud}.} \bibinfo{year}{2021}\natexlab{}.
\newblock \showarticletitle{Detecting and quantifying wash trading on
  decentralized cryptocurrency exchanges}. In
  \bibinfo{booktitle}{\emph{Proceedings of the Web Conference 2021}}.
  \bibinfo{pages}{23--32}.
\newblock


\bibitem[von Wachter et~al\mbox{.}(2022)]%
        {von2022nft}
\bibfield{author}{\bibinfo{person}{Victor von Wachter},
  \bibinfo{person}{Johannes~Rude Jensen}, \bibinfo{person}{Ferdinand Regner},
  {and} \bibinfo{person}{Omri Ross}.} \bibinfo{year}{2022}\natexlab{}.
\newblock \showarticletitle{NFT Wash Trading: Quantifying suspicious behaviour
  in NFT markets}.
\newblock \bibinfo{journal}{\emph{arXiv preprint arXiv:2202.03866}}
  (\bibinfo{year}{2022}).
\newblock


\bibitem[Wang et~al\mbox{.}(2021)]%
        {wang2021non}
\bibfield{author}{\bibinfo{person}{Qin Wang}, \bibinfo{person}{Rujia Li},
  \bibinfo{person}{Qi Wang}, {and} \bibinfo{person}{Shiping Chen}.}
  \bibinfo{year}{2021}\natexlab{}.
\newblock \showarticletitle{Non-fungible token (NFT): Overview, evaluation,
  opportunities and challenges}.
\newblock \bibinfo{journal}{\emph{arXiv preprint arXiv:2105.07447}}
  (\bibinfo{year}{2021}).
\newblock


\bibitem[Williams(2022)]%
        {meebits_wash}
\bibfield{author}{\bibinfo{person}{Chris Williams}.}
  \bibinfo{year}{2022}\natexlab{}.
\newblock \bibinfo{title}{A Meebit NFT Sold for \$49.5M in Ethereum—But
  There's a Catch}.
\newblock
  \bibinfo{howpublished}{\url{https://cryptobriefing.com/meebit-nft-sold-for-49-5m-ethereum-with-catch/}}.
\newblock


\end{thebibliography}

\clearpage
\appendix
\section{Appendix}
\subsection{Ethics}
Due to anonymous nature of crpytocurrency wallet addresses, this work does not raise any ethical issues.

\subsection{NFT transaction data collection approach}
\label{sec:NFT_tx_collect}
In Ethereum, there are two types of transactions: \textit{external} and \textit{internal} transactions. External transactions are initiated by external accounts. Transaction information such as receiver’s and sender’s addresses and transferred amount in ETH is recorded in the blockchain and readily available to anyone for reference (e.g., \textit{etherscan.io}).

Internal transactions are results of preceding transactions like byproducts of smart contract functionality, and they are not recorded.
In this case, the transaction only records the truth that a message sender made a call on smart contract functionality, and not detailed behaviors of what the contract did. 
Unfortunately, NFT functionalities are implemented in the manner of smart contracts, making it infeasible for the extraction of information on NFT transactions through simple referencing of transaction data.
Instead, we can use the \textit{event logs}, which are the result of LOG opcodes being executed in the EVM (Ethereum Virtual Machine) to gain a clear view of smart contract behaviors. Commonly, smart contracts record their activities as event logs, and NFT contracts are not exceptions.
For instance, when there is an ownership change for an NFT item, the details about the transmission such as owner address and token identifier are recorded as an event log. By collecting such events from NFT contracts, we can trace how NFT transactions are working in the NFT ecosystem.
The transfer log, which is a type of the event log, explains how many NFTs are transferred and who sends the tokens to whom consisting of a number of entities. 
From the transfer log, we can extract information of ownership change for an NFT item, including the token contract address, the sender’s and receiver’s accounts, and transferred token number.
A transfer log from one of the most popular NFT projects is described in Listing \ref{l:transfer_log} as an example.
Diving into the specifics of each of the log entities, the ‘address’ indicates the token contract address to launch the transfer. ‘Topics’ refers to a combination of four entities including the sender’s and receiver’s accounts and transferred token number. According to the NFT standard, one NFT is transferred at a time.

Most NFT buyers make payments in ETH or using several fungible tokens. However, the actual transaction price for an NFT sale is not included in a NFT transfer log. To figure out the transaction price for NFT sales, it is necessary to refer to additional data. 
More specifically, when the NFT buyer pays in ETH, the payment of sales is recorded in the external transaction. 
If the buyer paid in fungible tokens, a transfer log of fungible token explains how much tokens are transferred as payment for the NFT sales. 
Thus, if the transfer value of a transaction lists a value greater than 0 ETH, we consider the amount of transferred ETH as the actual price of NFT sales. If not, we look up transfer logs from any of fungible tokens transferred from an NFT buyer, and take the amount of transferred fungible tokens as the transaction price of NFT sales. Lastly, if there are no cryptocurrency payments from an NFT buyer for NFT transaction, the transaction is considered to just a transfer, not sale.

\subsection{Supplementary information of analysis}

\begin{table}[h]
\centering
\small
\begin{tabular}{c rrrr}
\toprule
\textbf{Year/Month} & \textbf{\# whales} & \textbf{\# dolphins} & \textbf{\# minnows} & \textbf{Total}\\
\midrule
21/01 & 6 & 590 & 5,368 & 5,964\\
21/02 & 13 & 1,237 & 11,255 & 12,505\\
21/03 & 20 & 1,977 & 17,973 & 19,970\\
21/04 & 25 & 2,465 & 22,412 & 24,902\\
21/05 & 38 & 3,755 & 34,137 & 37,930\\
21/06 & 45 & 4,456 & 40,601 & 45,012\\
21/07 & 61 & 5,995 & 54,499 & 60,555\\
21/08 & 117 & 11,578 & 105,257 & 116,952\\
21/09 & 179 & 17,745 & 161,318 & 179,242\\
21/10 & 248 & 24,530 & 223,001 & 247,779\\
21/11 & 292 & 28,936 & 263,047 & 292,275\\
21/12 & 362 & 35,804 & 325,396 & 361,662\\
22/01 & 403 & 39,868 & 362,443 & 402,714\\
22/02 & 430 & 42,593 & 387,204 & 430,227\\
\bottomrule
\end{tabular}
\centering
\caption{Number of each trader type on each month}
\label{t:num_of_trader}
\end{table}
\begin{figure}[h]
    \centering
    \includegraphics[width=0.9\linewidth]{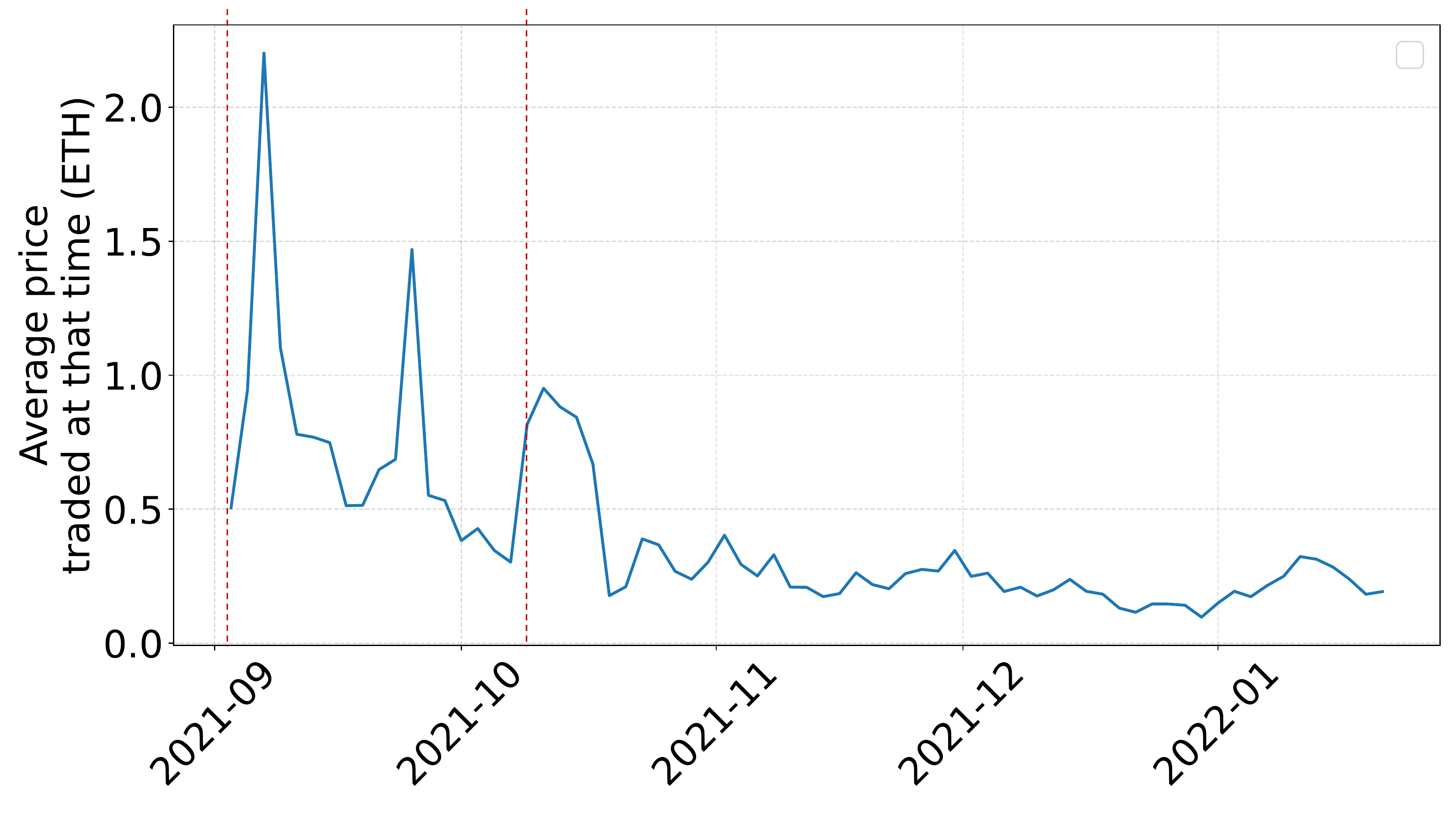}
    \caption{Average price of \textit{The n Project} NFTs traded at that time. Red line indicates when malicious behaviors occurred.}
    \label{fig:wash_price}
\end{figure}


\begin{table}[h]
\centering
\small
\begin{tabular}{l l r}
\toprule
\textbf{Rank} & \textbf{Collection} & \textbf{\# tokens} \\
\midrule
1 & Art Blocks & 9,548 \\
2 & The Sandbox & 9,549\\
3 & LOSTPOETS  & 4,766\\
4 & Meebits  & 2,744\\
5 & Hashmasks  & 2,671\\
6 & Mutant Ape Yacht Club & 1,621 \\
7 & RTFKT CLONE X + Murakami & 1,610\\
8 & CryptoPunks  & 1,599\\
9 & CyberKongz VX  & 1,558\\
10 & Punks Comic  & 1,528\\
\bottomrule
\end{tabular}
\vspace{0.3cm}
\centering
\caption{Top 10 collections and corresponding number of tokens held by whales}
\label{t:whale_collections}
\end{table}




\begin{table*}[h]
\centering
\small
\begin{tabular}{cc rrrr}
\toprule
\multirow{2}{*}{\textbf{t = YY/MM}} 
& \multirow{2}{*}{\textbf{\% (t-1) whale remains in (t) whale}}
& \multicolumn{4}{c}{\textbf{\% (t) whale comes from}}\\
\cline{3-6}
&  & \textbf{Whale} & \textbf{Dolphin} & \textbf{Minnow} & \textbf{No activity before} \\
\midrule
21/02 & 50\% &23\% (3) & 38\% (5) & 0\% (0) & 38\% (5) \\
21/03 & 62\% &40\% (8) & 25\% (5) & 0\% (0) & 35\% (7)\\
21/04 & 70\% &56\% (14) & 28\% (7) & 0\% (0) & 16\% (4)\\
21/05 & 76\% &50\% (19) & 37\% (14) & 3\% (1) & 11\% (4)\\
21/06 & 100\% &84\% (38) & 13\% (6) & 0\% (0) & 2\% (1)\\
21/07 & 80\% &59\% (36) & 21\% (13) & 3\% (2) & 16\% (10)\\
21/08 & 59\% &31\% (36) & 39\% (46) & 2\% (2) & 28\% (33)\\
21/09 & 86\% &56\% (101) & 31\% (56) & 2\% (4) & 10\% (18)\\
21/10 & 89\% &65\% (160) & 29\% (73) & 0\% (1) & 6\% (14)\\
21/11 & 93\% &79\% (230) & 17\% (49) & 1\% (3) & 3\% (10)\\
21/12 & 84\% &67\% (244) & 27\% (96) & 1\% (2) & 6\% (20)\\
22/01 & 88\% &79\% (319) & 15\% (60) & 0\% (1) & 6\% (23)\\
22/02 & 89\% &83\% (357) & 11\% (48) & 0\% (0) & 6\% (25)\\
\bottomrule
\end{tabular}
\caption{Percentage of each group where whale belonged to}
\label{t:across_users}
\end{table*}

\begin{table*}[h]
\centering
\small
\begin{tabular}{c cccccc}
\toprule
\multirow{2}{*}{\textbf{YY/MM}} 
& \multicolumn{2}{c}{\textbf{Whale}} & \multicolumn{2}{c}{\textbf{Dolphin}} & \multicolumn{2}{c}{\textbf{Minnow}} \\
\cline{2-7}
& \textbf{\# actives (\%)}  & \textbf{Avg \# transactions}  
& \textbf{\# actives (\%)} & \textbf{Avg \# transactions} 
& \textbf{\# actives (\%)} & \textbf{Avg \# transactions} \\
\midrule
21/01 & 6 (100\%) & 218 & 586 (100\%) & 28 & 5323 (100\%) & 7 \\
21/02 & 13 (100\%) & 138 & 1170 (95\%) & 19 & 8448 (76\%) & 5\\
21/03 & 19 (95\%) & 19 & 1475 (75\%) & 13 & 10599 (59\%) & 4\\
21/04 & 22 (88\%) & 30 & 1322 (54\%) & 12 & 8839 (40\%) & 4\\
21/05 & 36 (95\%) & 154 & 2872 (76\%) & 14 & 17009 (50\%) & 5\\
21/06 & 26 (58\%) & 90 & 2607 (59\%) & 18 & 14164 (35\%) & 6\\
21/07 & 53 (87\%) & 43 & 4373 (73\%) & 20 & 24765 (46\%) & 7\\
21/08 & 114 (97\%) & 123 & 11036 (95\%) & 27 & 75551 (72\%) & 7\\
21/09 & 148 (83\%) & 58 & 14874 (84\%) & 17 & 99257 (62\%) & 5\\
21/10 & 189 (76\%) & 41 & 18211 (74\%) & 10 & 120058 (54\%) & 3\\
21/11 & 196 (68\%) & 34 & 18583 (64\%) & 7 & 95574 (36\%) & 3\\
21/12 & 258 (72\%) & 46 & 24175 (68\%) & 8 & 132279 (41\%) & 3\\
22/01 & 292 (73\%) & 34 & 27361 (69\%) & 6 & 109103 (30\%) & 2\\
22/02 & 275 (64\%) & 26 & 21472 (50\%) & 4 & 78029 (20\%) & 2\\
\bottomrule
\end{tabular}
\caption{Transaction statistics of active users on each month}
\label{t:active_users}
\end{table*}

\begin{table*}[ht!]
\centering
\begin{tabular}{c|c|c|c|c|c}
\toprule
\textbf{\# transfer} & 1 & 2 & 3 & 4 & $\ge$ 5\\
\hline
\textbf{Avg profit} & 7848.8 & 8604.7 & 10750.1 & 9274.1 & 14675.6\\
\bottomrule
\end{tabular}
\vspace{0.3cm}
\centering
\caption{Average Profit from \textit{mint} according to number of transfer}
\label{t:mint_appendix1}
\end{table*}

\begin{table*}[ht!]
\centering
\begin{tabular}{c|c|c|c|c}
\toprule
\textbf{Time duration} & < 1 day & 1 day - 1 week & 1 week - 1 month & $\ge$ 1 month\\
\hline 
\textbf{Avg profit} & 3372.3 & 4285.7 & 4660.6 & 10585.3\\
\bottomrule
\end{tabular}
\vspace{0.3cm}
\centering
\caption{Average Profit from \textit{mint} according time duration before first \textit{sell}}
\label{t:mint_appendix2}
\end{table*}

\end{document}